\documentclass[epja,amsfonts,amsmath,twocolumn,superscriptaddress,nofootinbib]{revtex4}

\usepackage{amssymb,color,paralist,amsmath,epsfig}
\usepackage{graphicx}
\usepackage{comment}
\usepackage{amsfonts}
\usepackage{relsize}
\usepackage{nicefrac,esint}
\usepackage{float}

\usepackage[colorlinks,citecolor=blue,linktoc=all,linkcolor=cyan]{hyperref}

\newcommand{\beq}{\begin{equation}}
\newcommand{\eeq}{\end{equation}}

\newcommand{\ber}{\begin{eqnarray}}
\newcommand{\eer}{\end{eqnarray}}

\begin{document}

\title{Electromagnetic proton-neutron mass difference}

\author{Oleksandr Tomalak}
\affiliation{Institut f\"ur Kernphysik and PRISMA Cluster of Excellence, Johannes Gutenberg Universit\"at, Mainz, Germany}
\affiliation{Department of Physics and Astronomy, University of Kentucky, Lexington, KY 40506, USA}
\affiliation{Fermilab, Batavia, IL 60510, USA}

\date{\today}

\begin{abstract}
We discuss the Cottingham formula and evaluate the proton-neutron electromagnetic mass difference exploiting the state-of-the-art phenomenological input. We decompose individual contributions to the mass splitting into Born, inelastic and subtraction terms. We evaluate the subtraction-function contribution connecting the input based on experimental data with the operator product expansion matched to QCD which allows us to avoid model dependence and to reduce errors of this contribution. We evaluate inelastic and Born terms accounting for modern low-$Q^2$ data. 
\end{abstract}

\maketitle

Two isospin-violating effects inside nucleons, the difference between the up and down quark masses and electromagnetic interaction, result in the shift between the proton $M_p$ and neutron $M_n$ masses $\delta M_{p-n}$ \cite{Tanabashi:2018oca}:
\ber
\delta M_{p-n} = M_p - M_n = -1.29333217(42)~\mathrm{MeV}. \label{mass_difference_pdg}
\eer
It is well known that the QED contributions enter Eq. (\ref{mass_difference_pdg}) with a positive sign. The leading electromagnetic correction was related to the phenomenological input from the electron-proton scattering by Cottingham in Ref. \cite{Cottingham:1963zz} and investigated in detail together with ideas about the negative sign contributions in a historical review of Ref. \cite{Zee:1971df}. The origin of the negative sign due to the difference between up and down quark masses was pointed in Ref. \cite{Gasser:1974wd}, where authors evaluated as well the electromagnetic contribution: $\delta M^\gamma_{p-n} = 0.76\pm0.30~\mathrm{MeV}$. In Ref. \cite{Collins:1978hi}, the author has renormalized the Cottingham formula explicitly and pointed on the small correction from the high-energy counterterms. Recent studies of Ref. \cite{WalkerLoud:2012bg} accounted for the modern experimental data on the inelastic proton structure and have corrected the elastic contribution of Ref. \cite{Gasser:1974wd}. The new result $\delta M^\gamma_{p-n} = 1.30\pm0.47~\mathrm{MeV}$  \cite{WalkerLoud:2012bg} is within uncertainties of Refs. \cite{Gasser:1974wd,Gasser:2015dwa,Gasser:2020mzy}. However, the central values are quite different, which motivates to explore individual contributions to the Cottingham formula in detail. The electromagnetic effect was studied also in Refs. \cite{Thomas:2014dxa,Erben:2014hza,Endres:2015gda,Gasser:2015dwa}, while the QCD contribution to splitting was investigated in Refs. \cite{Beane:2006fk,deDivitiis:2011eh,Nasrallah:2012xz,Horsley:2012fw,Shanahan:2012wa}. Both corrections were evaluated on the lattice in Ref.~\cite{Blum:2010ym} with small lattice and relatively heavy pions as well with lower pion masses by BMW Collaboration~\cite{Borsanyi:2013lga,Borsanyi:2014jba}. The dispersive estimate of Ref. \cite{Thomas:2014dxa}: $\delta M^\gamma_{p-n} = 1.04\pm0.11~\mathrm{MeV}$, gave smaller uncertainty due to very optimistic assumptions about our knowledge of the subtraction function and of the isovector nucleon polarizability. The best lattice result with four nondegenerate quark flavors for the electromagnetic contribution is $\delta M^\gamma_{p-n} = 1.00\pm0.16~\mathrm{MeV}$ \cite{Borsanyi:2014jba}. It is in a good agreement with phenomenological estimates and has smaller error. The four-flavor result~\cite{Borsanyi:2014jba} is 1-2$\sigma$ smaller than the three-flavor calculation of Ref. \cite{Horsley:2015eaa}: $\delta M^\gamma_{p-n} = 1.71\pm0.30~\mathrm{MeV}$, and agrees within errors with Ref. \cite{Borsanyi:2013lga}: $\delta M^\gamma_{p-n} = 1.59\pm0.46~\mathrm{MeV}$ as well as with earlier three-flavor studies of Ref. \cite{Blum:2010ym} with the shift $\delta M^\gamma_{p-n} = 0.38 \pm 0.68~\mathrm{MeV}$ and with an exploratory work of Ref. \cite{Endres:2015gda}: $\delta M^\gamma_{p-n} = 0.53-0.84~\mathrm{MeV}$, which has demonstrated a new approach for such calculations exploiting an ensemble with the pion mass far from its physical value. To put constraints on the up-down quark mass difference, the lattice result of Ref. \cite{Borsanyi:2014jba} requires an independent cross check by dispersive calculation.

In this paper, we present the derivation of the Cottingham formula considering the decomposition into the Born, inelastic and subtraction contributions. We update the size of all terms relying on modern experimental input.

\begin{figure}[h]
\begin{center}
\includegraphics[width=.35\textwidth]{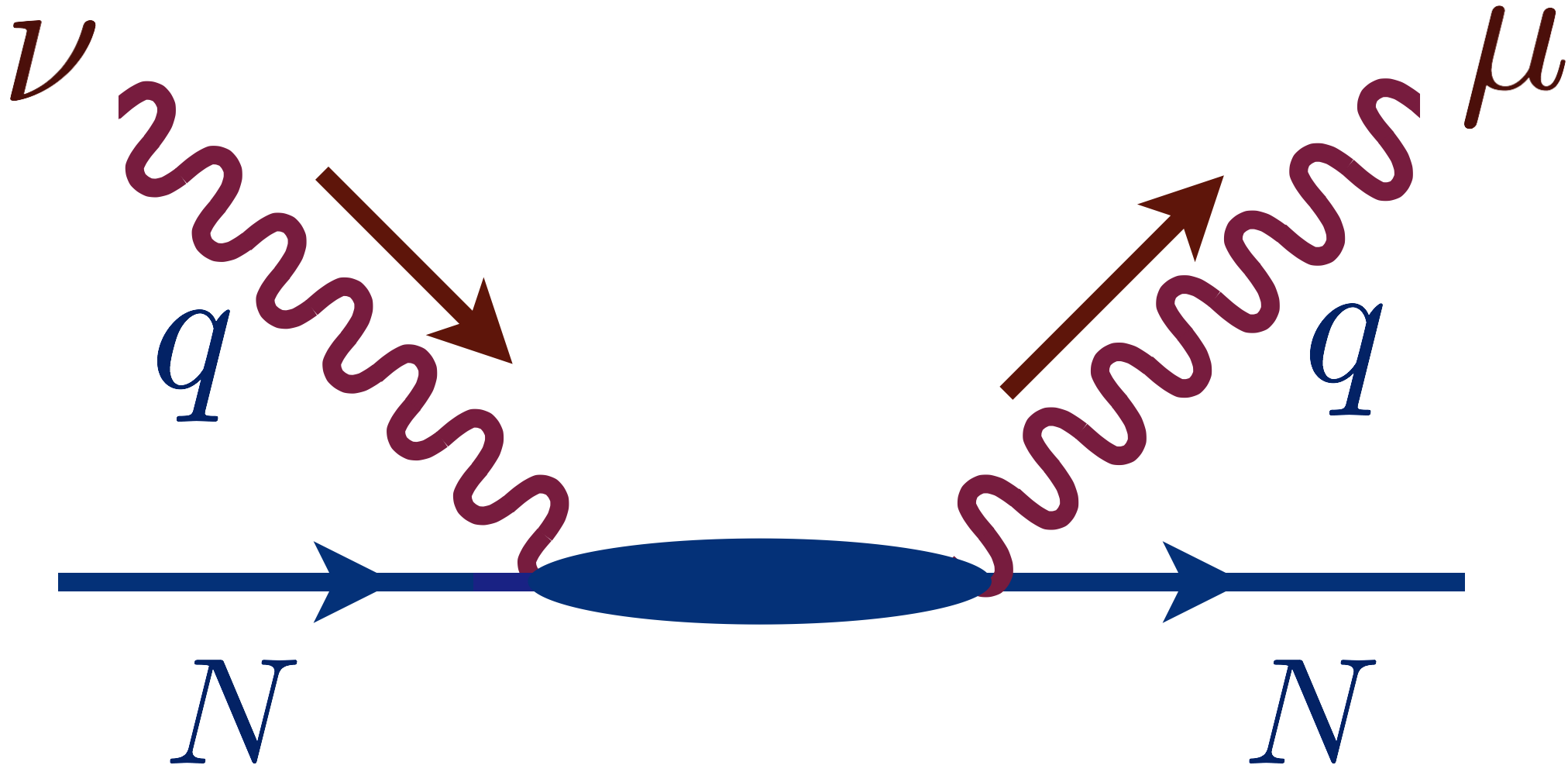}
\end{center}
\caption{Forward VVCS process.}
\label{VVCS_forward}
\end{figure}

The forward doubly virtual Compton scattering (VVCS) process on a nucleon (see Fig. \ref{VVCS_forward} for kinematics): $ \gamma^\ast\left(q\right)  + N\left(p\right) \to \gamma^\ast\left(q\right) + N\left(p\right) $, is described by the amplitude $ \mathrm{T} $. The latter can be expressed in terms of the forward VVCS tensor $ \mathrm{M}^{\mu \nu} $ as
\ber \label{forward_vvcs_helamp}
\mathrm{T} = \varepsilon_\nu\left(q\right) \varepsilon^\ast_\mu\left(q\right)  \bar{N}\left(p\right) (4 \pi  \mathrm{M}^{\mu \nu})  N\left(p\right),
\eer
where $ N, \bar{N} $ denote the nucleon spinors, $ \varepsilon_\nu, ~\varepsilon^\ast_\mu $ are the initial and final virtual photon polarization vectors. The nucleon is at rest in the laboratory frame, i.e., $p=(M,0)$, while the photon energy is given by $\nu_\gamma = \left( p \cdot q\right)/M$ and the virtuality is $Q^2 = -q^2$.

The nucleon self-energy correction is determined by the symmetric part of the forward VVCS tensor  $ \mathrm{M}^{\mu \nu}_S $:
\ber
\mathrm{M}^{\mu \nu}_S \hspace{-0.1cm}& = &\hspace{-0.1cm} \left( - g^{\mu\nu} + \frac{q^{\mu}q^{\nu}}{q^2}\right)
\mathrm{T}_1 (\nu_\gamma, Q^2) \nonumber \\
\hspace{-0.1cm}&+&\hspace{-0.1cm} \left(p^{\mu}-\frac{ \left(p\cdot
q \right)}{q^2}\,q^{\mu}\right) \left(p^{\nu}-\frac{\left( p\cdot
q \right)}{q^2}\, q^{\nu} \right) \frac{ \mathrm{T}_2 (\nu_\gamma, Q^2)}{M^2}, \nonumber \\  \label{forward_vvcs_tensor} 
\eer
with the unpolarized forward Compton amplitudes $ \mathrm{T}_1$ and $\mathrm{T}_2$, which enter Eq. (\ref{forward_vvcs_tensor}) in a gauge-invariant way, i.e., $ q_\mu \mathrm{M}^{\mu \nu} = q_\nu \mathrm{M}^{\mu \nu} = 0 $. The imaginary parts of the forward VVCS amplitudes $ \mathrm{T}_1$ and $\mathrm{T}_2 $ are related to the unpolarized proton structure functions $ \mathrm{F}_1$ and $ \mathrm{F}_2 $ by
\ber \label{imaginary_ampl}
\Im \mathrm{T}_1(\nu_\gamma, Q^2)  =  \frac{e^2}{4 M} \mathrm{F}_1(\nu_\gamma, Q^2), \\
\Im \mathrm{T}_2 (\nu_\gamma, Q^2)  =  \frac{e^2}{4 \nu_\gamma} \mathrm{F}_2(\nu_\gamma, Q^2),
\eer
where $e$ denotes the electric charge.

The real part of the even amplitude $ \mathrm{T}_1 $ is related to the imaginary part through the subtracted dispersion relation:
\ber \label{forward_VVCS_DR_T1}
\Re \mathrm{T}_1 (\nu_\gamma, Q^2) &  = &  \mathrm{T}^{\mathrm{\mathrm{subt}}}_1 (0, Q^2) + \Re \mathrm{T}^{\mathrm{Born}}_1 ( \nu_\gamma, Q^2 ) \nonumber \\
&+&  \frac{e^2 \nu_\gamma^2}{2\pi}  \fint \limits^{~~ \infty}_{\nu^{\mathrm{inel}}_{\mathrm{thr}} } \frac{ \mathrm{F}_1\left(\nu',Q^2\right) \mathrm{d} \nu'}{M \nu' \left( \nu'^2 - \nu^2_\gamma\right)}, \label{T1_amplitude}
\eer
with the pion-nucleon production threshold: $ \nu^{\mathrm{inel}}_{\mathrm{thr}} = m_\pi +  \left( m_\pi^2 + Q^2 \right) / \left( 2 M \right)$, where $ m_{\pi} $ denotes the pion mass, $  \mathrm{T}^{\mathrm{\mathrm{subt}}}_1 (0, Q^2 ) $ is the subtraction function at zero photon energy $ \nu_\gamma = 0$, and $\mathrm{F}_1$ contains only the inelastic contributions since we have separated the Born piece \cite{Birse:2012eb}. The real part of the unpolarized amplitude $  \mathrm{T}_2 $ can be obtained from the unsubtracted dispersion relation:
\ber \label{forward_VVCS_DR_T2}
\Re \mathrm{T}_2 (\nu_\gamma, Q^2) \hspace{-0.12cm} &=&  \hspace{-0.12cm} \Re \mathrm{T}^{\mathrm{Born}}_2 ( \nu_\gamma, Q^2 ) +  \frac{e^2}{2\pi}   \fint \limits^{~~ \infty}_{\nu^{\mathrm{inel}}_{\mathrm{thr}} } \frac{\mathrm{F}_2 \left(\nu', Q^2\right)  \mathrm{d} \nu'}{\nu'^2 - \nu^2_\gamma}. \nonumber \\ \label{T2_amplitude}
\eer

\begin{figure}[h]
\begin{center}
\includegraphics[width=.35\textwidth]{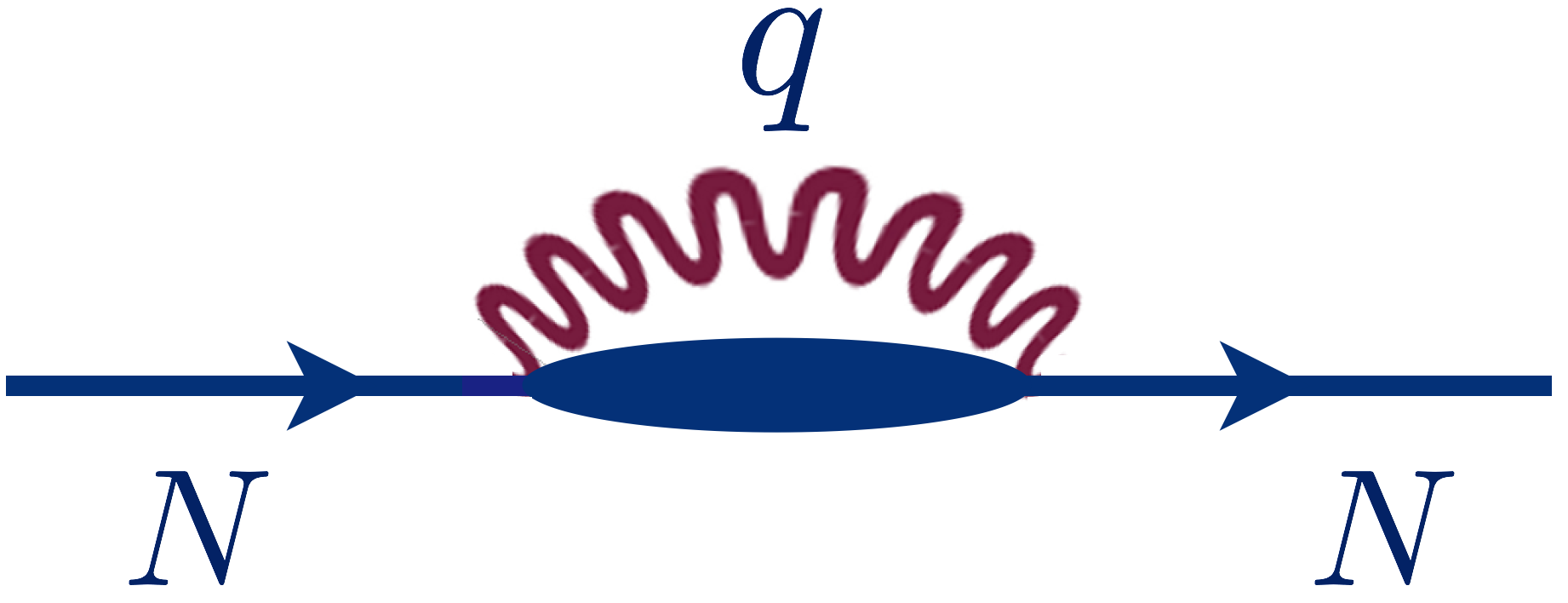}
\end{center}
\caption{Nucleon self-energy correction.}
\label{self_energy}
\end{figure}

The self-energy electromagnetic correction to the nucleon propagator, see Fig. \ref{self_energy}, is given by
\ber
\mathrm{S} - \mathrm{S}_0 \hspace{-0.05cm} &=& \hspace{-0.05cm} \mathrm{S} _0 \left( \frac{1}{2} \int \frac{\mathrm{d}^4 q}{\left(2\pi \right)^4}  \frac{- i g_{\mu \nu}}{q^2} 4 \pi i \mathrm{M}^{\mu \nu} \right) \mathrm{S}, \label{propagators_correction}
\eer
with the full propagator $\mathrm{S} $ and the free propagator $\mathrm{S} _0$:
\ber
\mathrm{S}_0 =  \frac{i}{\hat{p} - M},
\eer
where $ \hat{a} = \gamma^\mu a_\mu$. Multiplying Eq. (\ref{propagators_correction}) by $\mathrm{S}_0^{-1}$ from the left and by $\mathrm{S}^{-1}$ from the right, we obtain:
\ber
\mathrm{S}_0^{-1} - \mathrm{S}^{-1} =  \int \frac{i \mathrm{d}^4 q}{\left(2\pi \right)^3}  \frac{\mathrm{M}^{\mu}_\mu}{q^2},
\eer
resulting into the electromagnetic mass shift $\delta M^\gamma$ \cite{Cottingham:1963zz}:
\ber
\delta M^\gamma =   \int \frac{i \mathrm{d}^4 q}{\left(2\pi \right)^3}  \frac{\mathrm{M}^{\mu}_\mu}{q^2}.
\eer
To relate it to the experimental input, we perform the Wick rotation first: $q_0 \to i \nu_\gamma $, and introduce the space-like virtuality $Q^2 = - q^2$. The mass shift is given by
\ber
\delta M^\gamma =   \int \frac{\mathrm{d} \nu \mathrm{d}^3 q}{\left(2\pi \right)^3}  \frac{\mathrm{M}^{\mu}_\mu}{Q^2} = \int \frac{\mathrm{d} \nu_\gamma \mathrm{d}Q^2}{\left(2\pi \right)^2} \frac{\sqrt{Q^2 - \nu_\gamma^2}\mathrm{M}^{\mu}_\mu}{Q^2}.
\eer
Changing the integration order and accounting for the crossing properties of the Compton scattering, the Cottingham formula \cite{Cottingham:1963zz} gives:
\ber
\delta M^\gamma =  \int \limits_0^{~\infty} \frac{\mathrm{d} Q^2}{\left(2\pi \right)^2}\int \limits^{~1}_{0}\frac{\sqrt{1 - \tilde{\tau}} \mathrm{d} \tilde{\tau} }{\sqrt{\tilde{\tau}}} \mathrm{M}^{\mu}_\mu,
\eer
with $\tilde{\tau} = \nu_\gamma^2/Q^2$ and the trace of the forward VVCS tensor:
\ber
\mathrm{M}^{\mu}_\mu = - 3 \mathrm{T}_1 (i \nu_\gamma, Q^2) + \left( 1 - \tilde{\tau} \right) \mathrm{T}_2 (i \nu_\gamma, Q^2) .
\eer

Following the decomposition of the forward VVCS amplitudes of Eqs. (\ref{T1_amplitude}) and (\ref{T2_amplitude}), we introduce the Born contribution $\delta \mathrm{M}^{\mathrm{Born}}$, the inelastic correction $\delta \mathrm{M}^{\mathrm{inel}}$ and the subtraction term $\delta \mathrm{M}^{\mathrm{subt}}$:
\ber
\delta M^\gamma = \delta M^{\mathrm{Born}} +\delta M^{\mathrm{inel}} + \delta M^{\mathrm{subt}}.
\eer

\begin{figure}[h]
\begin{center}
\includegraphics[width=.5\textwidth]{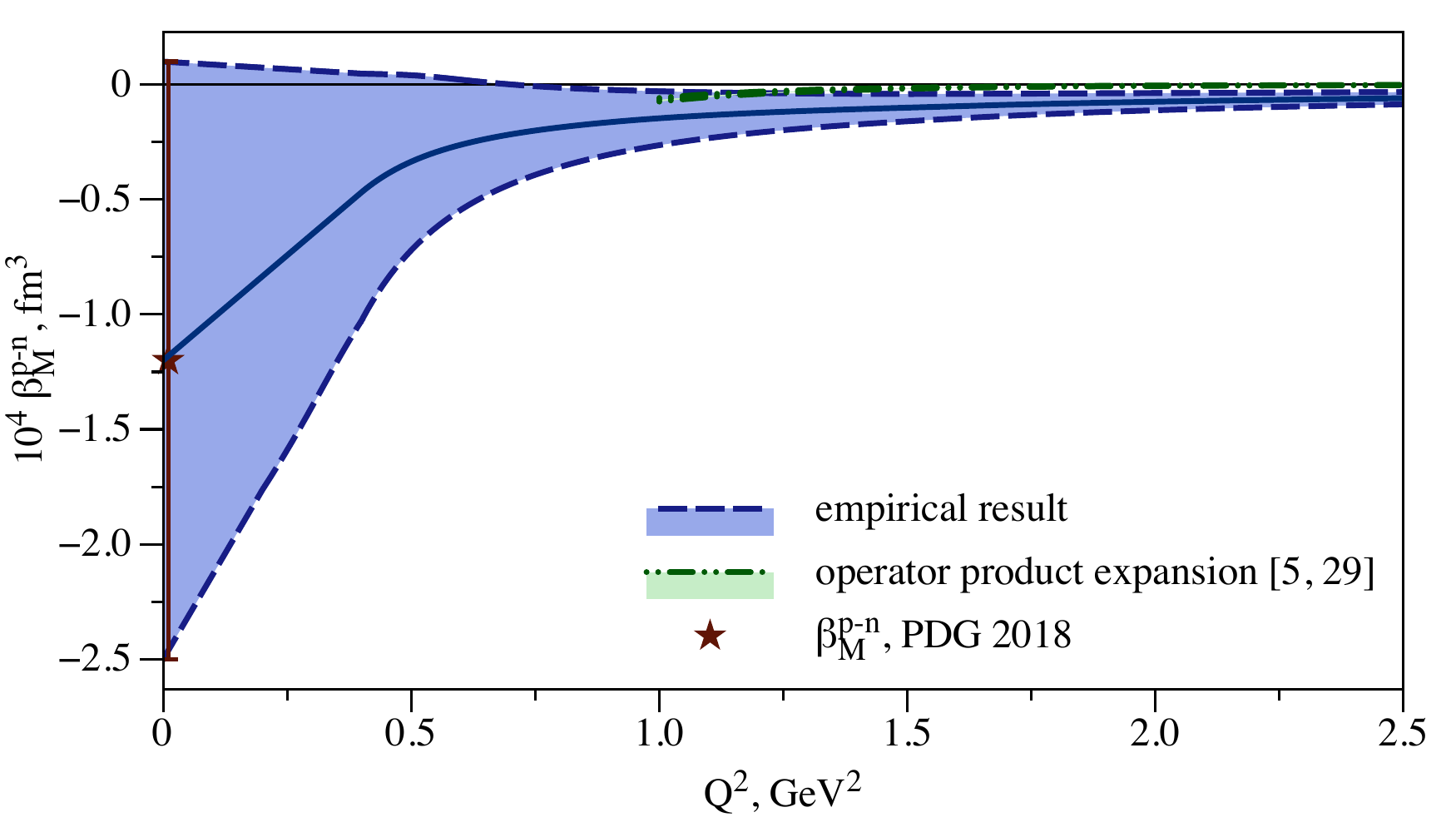}
\end{center}
\caption{The effective isovector magnetic polarizabiltity $\beta^{p-n}_M (Q^2) =\mathrm{T}_{1,p-n}^{\mathrm{subt}} \left( 0,Q^2\right)/Q^2$ based on Refs.~\cite{Tomalak:2015hva,Tomalak:2018uhr,Pilkuhn:1973wq,Donnachie:2004pi,Bosted:2007xd,Christy:2007ve} vs \cite{Collins:1978hi,Hill:2016bjv}. }
\label{beta_subtraction}
\end{figure}

Exploiting the integral:
\ber
\int \limits^{~1}_{0}\frac{\sqrt{1 - \tilde{\tau}} \mathrm{d} \tilde{\tau} }{ \sqrt{\tilde{\tau}}} = \frac{\pi}{2}, \label{int1}
\eer
the contribution of the subtraction function $\mathrm{T}_{1,p-n}^{\mathrm{subt}} \left( 0,Q^2\right)$ to the proton-neutron mass difference $\delta \mathrm{M}^{\mathrm{subt}}_{p-n}$ can be easily expressed as \cite{Gasser:1974wd,Gasser:2015dwa,WalkerLoud:2012bg}
\ber
\delta M^{\mathrm{subt}}_{p-n} =  - \frac{3}{8 \pi} \int \limits_0^{~\infty} \mathrm{d} Q^2 \mathrm{T}_{1,p-n}^{\mathrm{subt}} \left( 0,Q^2\right).
\eer
Instead of evaluating the isovector magnetic polarizability from the derivative of the longitudinal to transverse cross sections ratio at origin relying on four data points at relatively large virtuality $Q^2 \gtrsim 0.75~\mathrm{GeV}^2$ \cite{Sibirtsev:2010zg,Sibirtsev:2013cga}  with assumption of energy independence and isospin symmetry \cite{Gasser:2015dwa}, we take the difference between the proton $\beta^p_M$ and neutron $\beta^n_M$ magnetic polarizabilities: 
\ber
\beta^{p-n}_M = \beta^p_M - \beta^n_M = \left(-1.2 \pm 1.3 \right)\times10^{-4}~\mathrm{fm}^3,
\eer
from p.d.g. \cite{Tanabashi:2018oca} and estimate the subtraction function at higher $Q^2$ evaluating the unsubtracted dispersion relation for the amplitude free from the Regge high-energy behavior (Refs.~\cite{Tomalak:2015hva,Tomalak:2018uhr}) with an input from Refs. \cite{Pilkuhn:1973wq,Donnachie:2004pi,Bosted:2007xd,Christy:2007ve}. We estimate the uncertainty of the proton structure functions at $3~\%$ level, double the error for the neutron structure functions and assign a $30~\%$ uncertainty to a Reggeon pole residue \cite{Gasser:2015dwa}. We connect the experimental isovector magnetic polarizability, with the p.d.g. value at zero virtuality $ \beta^{p-n}_M \left( 0 \right)= \beta^{p-n}_M$, and higher-$Q^2$ region on the level of $\beta^{p-n}_M \left( Q^2 \right)=\mathrm{T}_{1,p-n}^{\mathrm{subt}} \left( 0,Q^2\right)/Q^2$; see Fig.~\ref{beta_subtraction} for details.
The subtraction term contributes:
\ber
\delta M^{\mathrm{subt}}_{p-n} = 0.33 \pm 0.30~\mathrm{MeV},
\eer
where we have chosen the upper integration limit at relatively low $Q^2 = 1~\mathrm{GeV}^2$ and have exploited the pQCD result based on the operator product expansion~\cite{Hill:2016bjv} above which gives a rough estimate above and in the region of hadronic physics.\footnote{Without inputs from Ref.~\cite{Hill:2016bjv}, the subtraction term contributes $\delta M^{\mathrm{subt}}_{p-n} = 0.54 \pm 0.46~\mathrm{MeV}$. } According to Ref.~\cite{Hill:2016bjv}, the resulting Compton amplitude $\mathrm{T}_{1,p-n} (0,~Q^2)$ is much smaller than the Born contribution. Consequently, the subtraction function in QCD region can be well approximated as $\mathrm{T}_{1,p-n}^{\mathrm{subt}} (0,~Q^2) \approx - \mathrm{T}_{1,p-n}^{\mathrm{Born}} (0,~Q^2)$ which has the same sign but is smaller up to an order of magnitude than any phenomenological estimate~\cite{WalkerLoud:2012bg,Thomas:2014dxa}, and contributes $0.06~\mathrm{MeV}$ to mass difference from this region. We have added uncertainties of the subtraction-function contribution: $0.28~\mathrm{MeV}$, and due to the variation of the upper integration limit over the range $1-1.5~\mathrm{GeV}^2$: $0.1~\mathrm{MeV}$, in quadrature. Our result is within uncertainties of previous estimates: $\delta M^{\mathrm{subt}}_{p-n} = 0.47 \pm 0.47~\mathrm{MeV}$~\cite{WalkerLoud:2012bg}, $\delta M^{\mathrm{subt}}_{p-n} = 0.21 \pm 0.11~\mathrm{MeV}$~\cite{WalkerLoud:2012bg,Thomas:2014dxa}. Our central value is determined by the isovector nucleon magnetic polarizability~\cite{Weller:2009zza,Sokhoyan:2016yrc,Downie:2016ace,Annand:2013ace}, and our error is smaller than the previous data-driven evaluations. In order to compete with the lattice calculation of Ref. \cite{Borsanyi:2014jba}, besides the necessary improvement in the structure functions in the resonance and DIS regions, the uncertainty on the isovector magnetic polarizability has to be reduced to $\left(0.4-0.5\right)\times10^{-4}~\mathrm{fm}^3$, at least. Moreover, additional studies within the framework of low-energy effective field theories \cite{Alarcon:2013cba,Lensky:2017bwi} could shed more light on the most uncertain low-$Q^2$ region.

We obtain the Born contribution substituting the corresponding unpolarized Compton amplitudes $\mathrm{T}_1^{\mathrm{Born}}$ and $\mathrm{T}_2^{\mathrm{Born}}$:
\ber
\mathrm{T}_1^{\mathrm{Born}} \left( \tilde{\tau}, Q^2 \right)  \hspace{-0.05cm} & = & \hspace{-0.05cm}\frac{\alpha}{M} \left(  \frac{ \mathrm{G}^2_\mathrm{M}\left(Q^2\right)}{ 1 - \frac{\tilde{\tau}}{\tau_\mathrm{P}}- i \varepsilon} - \mathrm{F}_\mathrm{D}^2 \left(Q^2\right) \right), \label{t1_Born} \\
\mathrm{T}_2^{\mathrm{Born}} \left( \tilde{\tau}, Q^2 \right) \hspace{-0.05cm} & = & \hspace{-0.05cm}\frac{\alpha}{M}  \frac{  \mathrm{G}_\mathrm{E}^2\left(Q^2\right) + \tau_\mathrm{P} \mathrm{G}_\mathrm{M}^2\left(Q^2\right) }{\tau_\mathrm{P} \left( 1 + \tau_\mathrm{P} \right)  \left(1 - \frac{\tilde{\tau}}{\tau_\mathrm{P}}- i \varepsilon \right)},
\label{t2_Born}  
\eer
with the Dirac ($ \mathrm{F}_\mathrm{D}$), Sachs electric ($ \mathrm{G}_\mathrm{E} $) and magnetic ($ \mathrm{G}_\mathrm{M} $) form factors, the electromagnetic coupling constant $\alpha \equiv e^2 / \left( 4 \pi \right)$ and the notation $\tau_\mathrm{P} = Q^2/ \left( 4 M^2\right)$. Introducing the additional notation $\rho \left( \tau \right)$:
\ber
\rho \left( \tau \right) = 2 \left(\tau - \sqrt{\tau ( 1 + \tau )} \right),
\eer
and exploiting the integral:
\ber
\int \limits^{~1}_{0}\frac{\sqrt{1 - \tilde{\tau}} \mathrm{d} \tilde{\tau} }{ \sqrt{\tilde{\tau}}\left(1 + \frac{\tilde{\tau}}{\tau}\right)} = - \frac{ \pi}{2} \rho \left( \tau \right), \label{int2}
\eer
we express the Born contribution to the proton-neutron mass difference $ \delta \mathrm{M}^{\mathrm{Born}}_{p-n}$ as
\ber
\delta M^{\mathrm{Born}}_{p-n} &=& \frac{3 \alpha}{8 \pi M}  \int \limits_0^{~\infty} \mathrm{d} Q^2 \left( \mathrm{F}_\mathrm{D}^2 \left(Q^2\right) +  \rho \left( \tau_\mathrm{P} \right)  \mathrm{G}_\mathrm{M}^2 \left(Q^2\right) \right)\nonumber \\
&-& \frac{\alpha}{8 \pi M}  \int \limits_0^{~\infty} \mathrm{d} Q^2 \left( 1+  \frac{ 1+ \tau_\mathrm{P}}{\tau_\mathrm{P} } \rho \left( \tau_\mathrm{P} \right)\right) \nonumber \\
&\times& \frac{  \mathrm{G}_\mathrm{E}^2\left(Q^2\right) + \tau_\mathrm{P} \mathrm{G}_\mathrm{M}^2\left(Q^2\right) }{1 + \tau_\mathrm{P}}, \label{Born_result} 
\eer
implying that the difference between proton and neutron contributions should be taken. For the numerical evaluation, we take the up-to-date proton form factors with uncertainties from Refs. \cite{Bernauer:2010wm,Bernauer:2013tpr} and the neutron form factors from Refs. \cite{Kubon:2001rj,Warren:2003ma,Kelly:2004hm,Punjabi:2015bba,Ye:2017gyb}. For the neutron, we obtain the central value averaging over the form factor parameterizations and estimate the uncertainty as a difference between the largest and smallest results. The resulting Born contribution is given by
\ber
\delta M^{\mathrm{Born}}_{p-n} = 0.74 \pm 0.01 ~\mathrm{MeV}.
\eer
The corrections to the proton mass $\delta M^{\mathrm{Born}}_{p}$:
\ber
\delta M^{\mathrm{Born}}_{p} = 0.53\pm0.01~\mathrm{MeV},
\eer
and neutron mass  $\delta M^{\mathrm{Born}}_{n}$:
\ber
\delta M^{\mathrm{Born}}_{n} = -0.21\pm0.01~\mathrm{MeV},
\eer
have an opposite sign enhancing the electromagnetic mass difference.\footnote{The proton and neutron mass corrections based on the state-of-the-art fits from Ref.~\cite{Borah:2020gte} are in agreement with our results and have smaller uncertainties.} Note that the analytical expression of Eq. (\ref{Born_result}) has no analogous in Ref. \cite{WalkerLoud:2012bg}; the difference is in the $\mathrm{G}^2_M$ contribution to the subtraction term \cite{Birse:2012eb,WalkerLoud:2012bg}. Apparently, this mismatch was accounted in the numerical evaluation, since the result of Ref. \cite{WalkerLoud:2012bg} for the whole elastic contribution: $0.77~\mathrm{MeV}$, is quite close to ours. 

With the same integrals of Eqs. (\ref{int1}) and (\ref{int2}), the inelastic contribution is expressed in terms of the unpolarized structure functions as
\ber
\delta M^{\mathrm{inel}}_{p-n}  \hspace{-0.16cm}  &=&  \hspace{-0.16cm}  - \frac{\alpha}{4 \pi}   \int \limits_0^{~\infty} \mathrm{d} Q^2 \int \limits^{\infty}_{\nu^\mathrm{inel}_\mathrm{thr}} \frac{\mathrm{d} \nu_\gamma}{\nu_\gamma} \left \{ \rho \left( \tilde{\tau} \right) \frac{\mathrm{F}_2 \left(\nu_\gamma, Q^2 \right)}{\nu_\gamma} \right. \nonumber \\
 \hspace{-0.2cm}  &-&  \hspace{-0.2cm} \left.  \left( 1 + \rho \left( \tilde{\tau} \right)\right) \left( \frac{3 \mathrm{F}_1 \left(\nu_\gamma, Q^2 \right)}{M} - \frac{ \nu_\gamma \mathrm{F}_2 \left(\nu_\gamma, Q^2 \right)}{Q^2} \right)  \right \}, \nonumber \\
\eer
implying that the difference between proton and neutron contributions should be taken. This is exactly the result of Refs. \cite{Gasser:1974wd,Gasser:2015dwa,WalkerLoud:2012bg}. We account for the inelastic correction relying on structure function from Refs. \cite{Donnachie:2004pi,Bosted:2007xd,Christy:2007ve}:
\ber
\delta M^\mathrm{inel}_{p-n} = 0.034 \pm 0.010 ~\mathrm{MeV},
\eer
where only the error due to the variation of upper integration region in the range $1.5~\mathrm{GeV}^2 < Q^2 < 2.5~\mathrm{GeV}^2$ included and the central value corresponds with the upper limit $Q^2 \le 2~\mathrm{GeV}^2$. Neglecting the counterterms contribution~\cite{Collins:1978hi,WalkerLoud:2012bg}, the resulting mass difference $\delta M^\gamma_{p-n}$ is given by
\ber
\delta M^\gamma_{p-n} = 1.09 \pm 0.3 ~\mathrm{MeV}.
\eer
Note that a conservative assignment of $100~\%$ error to inelastic contribution\footnote{A naive assignment of 3-5~\% error to nucleon structure functions gives an uncertainty estimate comparable to the inelastic contribution. A proper error estimate calls for reanalysis of nucleon structure functions taking correlations into account.} as well as increase in uncertainty from the Born term will not change the error of the resulting mass difference within significant digits.

We have presented the Cottingham formula in terms of the phenomenological input. We have updated the Born correction and estimated the subtraction term based on the experimental input. Our total result is within errors of the previous estimates \cite{Gasser:1974wd,WalkerLoud:2012bg,Thomas:2014dxa} due to the large uncertainty of the correction from the subtraction function. However, the knowledge of the Born contribution and the subtraction term is improved. Precise studies of the proton and neutron magnetic polarizabilities, inelastic structure functions and Regge trajectories will be able to improve the dispersive evaluation further. 

We acknowledge Vadim Lensky and Marc Vanderhaeghen for useful discussion. This work was supported  in part by the Deutsche Forschungsgemeinschaft (DFG) through Collaborative Research Center ``The Low-Energy Frontier of the Standard Model'' (SFB 1044), in part by a NIST precision measurement grant and by the U. S. Department of Energy, Office of Science, Office of High Energy Physics, under Award No. DE-SC0019095, and in part by the Visiting Scholars Award Program of the Universities Research Association.

\end{document}